\documentclass[aps,prd,a4paper,twocolumn,showpacs,nofootinbib]{revtex4}
\usepackage{amssymb,amsmath,latexsym,mathrsfs}
\usepackage{graphicx}
\usepackage{epsfig}
\usepackage{varioref,xr-hyper}
\usepackage{color}

\usepackage[hypertex,draft=false,verbose=false,debug=false
            hyperindex=true,hypertexnames=true]{hyperref}

\newcommand{\hide}[1]{{}}

\newcommand{\be}{\begin{equation}}
\newcommand{\ee}{\end{equation}}
\newcommand{\bea}{\begin{eqnarray}}
\newcommand{\eea}{\end{eqnarray}}

\newcommand{\begm}{\begin{pmatrix}}
\newcommand{\enm}{\end{pmatrix}}

\def\lsim{\;\raise 0.4ex\hbox{$<$}\kern -0.8em\lower 0.62 ex\hbox{$\sim$}\;}
\def\gsim{\;\raise 0.4ex\hbox{$>$}\kern -0.7em\lower 0.62 ex\hbox{$\sim$}\;}

\begin{document}

\title{Cosmic Microwave Weak
 lensing data as a test for the dark universe}

\author{Erminia Calabrese}
\affiliation{Physics Department, Universita' di Roma ``La Sapienza'',
  Ple Aldo Moro 2, 00185, Rome, Italy}
\author{An\v{z}e Slosar}
\email{anze@berkeley.edu}
\affiliation{Berkeley Center for Cosmological Physics Physics
  Department, University of California, Berkeley CA 94720}

\author{Alessandro Melchiorri}
\email{alessandro.melchiorri@roma1.infn.it}
\affiliation{Physics Department and INFN, Universita' di Roma ``La
  Sapienza'', Ple Aldo Moro 2, 00185, Rome, Italy}

\author{George F. Smoot}
\email{gfsmoot@lbl.gov}
\affiliation{Lawrence Berkeley National Laboratory and
Berkeley Center for Cosmological Physics Physics
  Department, University of California, Berkeley CA 94720}

\author{Oliver Zahn}
\email{zahn@berkeley.edu}
\affiliation{Lawrence Berkeley National Laboratory and
Berkeley Center for Cosmological Physics Physics
  Department, University of California, Berkeley CA 94720}

\begin{abstract}
\noindent

Combined analyses of WMAP 3-year and ACBAR Cosmic Microwave
Anisotropies angular power spectra have presented evidence 
for gravitational lensing at $>3$-$\sigma$
level. This signal could provide a relevant test for cosmology.
After evaluating and confirming the statistical significance of the
 detection in light of the new WMAP 5-year data, 
we constrain a new parameter $A_L$ that scales the lensing
potential such that $A_L=0$ corresponds to unlensed while $A_L=1$ is
the expected lensed result. We find from WMAP5+ACBAR a $2.5$-$\sigma$ 
indication for a lensing contribution larger than expected, with 
$A_L=3.1_{-1.5}^{+1.8}$ at $95 \%$ c.l..  The result is stable under the assumption of
different templates for an additional Sunyaev-Zel'dovich foreground
component or the inclusion of an extra background 
of cosmic strings. We find negligible correlation with other
cosmological parameters as, for example, the energy density in massive neutrinos.
While unknown systematics may be present, dark energy or 
modified gravity models could be responsible for the over-smoothness of the
power spectrum. Near future data, most notably from the Planck
satellite mission, will scrutinize this interesting possibility.
\end{abstract}


\maketitle

\section{Introduction}

Results from the last decade of Cosmic Microwave Background (hereafter CMB)
anisotropy observations have lead to a revolution in the field
of cosmology (see e.g.  ~\cite{boom}, \cite{maxima}, \cite{cbi}, 
\cite{wmap3cosm}, \cite{wmap5cosm}, \cite{wmap5komatsu}). Many fundamental
parameters of the cosmological model have now been measured with high
accuracy. Moreover, since the standard cosmological model of structure
formation, based on dark matter, inflation and a cosmological
constant, is in reasonable agreement with the current observations,
CMB anisotropies are now considered as a cosmological laboratory where
fundamental theories can be tested at scales and energies not
achievable on earth.

One crucial test concerns the nature of the dark
energy component and the validity of General Relativity (GR,
hereafter). The simple fact that supernovae type Ia observations are
in agreement with an accelerating universe, which is puzzling in
several theoretical respects, calls for the deepest possible
investigation of dark energy and for a continuous test of GR.

CMB anisotropies are mainly formed at redshift $z \sim 1000$ when
either dark energy or modifications to GR appear to be
negligible.  However, while CMB photons travel to us, they are
affected and distorted by other, low redshift, mechanisms, that could
help in understanding the nature of the accelerating universe.

The so-called late Integrated Sachs-Wolfe effect, for example,
generated by the time-variation of the gravitational potential field
along the CMB photon's line of sight in dark energy dominated
universes, has already been detected by more than five groups by cross
correlating galaxy surveys with anisotropies at very large angular
scales (see e.g. \cite{iswpap}).  
While the statistical significance of the effect is still
under $5 \sigma$, the detection represents a crucial test for dark
energy \cite{isw}.

On scales of ten arcminutes and smaller, the interaction of the CMB 
photons with the local universe starts to be dominant with second
order anisotropies arising from weak lensing or scattering of the CMB photons off 
ionized gas  in clusters and large scale structure (Sunyaev-Zel'dovich - SZ effect).

Weak lensing of CMB anisotropies could provide useful cosmological
information. Gravitational lensing cannot change the gross
distribution of primary CMB anisotropies, but it may redistribute
power and smooth the acoustic oscillations in the CMB power spectrum
(see e.g. \cite{lensteo}). Only in the tails of Silk damping
(\cite{silk}, at $\ell
\gsim 3000$) the lensing contribution start to change the power
spectrum significantly.  Higher signal-to-noise can be achieved by
correlating power in different directions on the sky, effectively
using the four-point function signature imprinted by lensing to
reconstruct the line-of-sight integrated matter
distribution\footnote{This type of estimator has recently been used to
  find evidence of order $3-\sigma$ in the WMAP data \cite{smith07,
    hirata08} in cross-correlation with galaxy surveys.}.
 
The strength of the weak lensing smoothing is related to the growth
rate and amplitude of the dark matter fluctuations.  Since both dark
energy or modified gravity significantly affects these perturbations,
a measurement of the CMB lensing, through its high-$\ell$ smoothing,
can in principle be a useful cosmological test (see
e.g. \cite{Acquaviva:2004fv}).

The recent claim made by the ACBAR collaboration (\cite{acbar}) for a
detection of weak lensing, based solely on smoothing of the angular
power spectrum, opens the opportunity for this kind of analysis.  To
first order, lensing causes the primordial peak structure to be less
pronounced, as gravitational potential fluctuations on large scales
mix the various scales in the primordial CMB power. Based on the
effect on the power spectrum, the ACBAR collaboration has reported a
$\Delta \chi^2 =9.46$ between the lensed and unlensed best fits to the
WMAP+ACBAR data, which translates into a $\ge 3 \sigma$ detection of
CMB lensing.

In this paper we further analyze this result and we study the possible
cosmological implications. In the next section we phenomenologically
uncouple weak lensing from primary anisotropies by introducing a new
parameter $A_L$ that scales the gravitational potential in a way such
that $A_L=1$ corresponds to the expected weak lensing scenario.  We
then constrain this parameter with current CMB data, we evaluate the
consistency with $A_L=1$, the correlation with other parameters and
with other systematics such as SZ. We will report a $\sim 2
\sigma$ preference for values of $A_L > 1$.  We will then discuss some
possible cosmological mechanisms that can increase the CMB smoothing,
namely an extra background of cosmic strings and modified gravity.

\section{Analysis Method}

Weak lensing of the CMB anisotropies enters as a convolution of the
unlensed temperature spectrum $C_\ell$ with the lensing potential
power spectrum $C_\ell^{\Psi}$ (see \cite{lensteo}). This
convolution serves to smooth out the main peaks in the unlensed
spectrum, which is the main qualitative effect on the power
spectrum on scales larger than the ACBAR beam, or $6 '$.

The weak lensing parameter is defined as a fudge 
scaling parameter affecting the lensing potential power spectrum:
\begin{equation}
  \label{eq:wel}
  C_\ell^{\Psi} \rightarrow A_L C_\ell^{\Psi}.
\end{equation}

\begin{figure}
\centerline{\includegraphics[width=8cm]{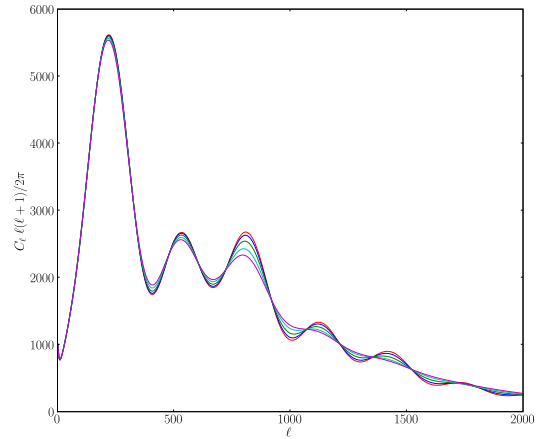}}
\caption{\label{fig:eff}This figure shows the effect of varying $A_L$ 
parameter. The curves with increasingly smoothed peak structure 
correspond to values of $A_L$ of $0$,$1$,$3$,$6$,$9$.}
\end{figure}

In other words, parameter $A_L$ effectively multiplies the matter
power lensing the CMB by a known factor. $A_L=0$ is therefore
equivalent to a theory that ignores lensing of the CMB, while $A_L=1$
gives the standard lensed theory. Since at the scales of interest
the main effect of lensing is purely to smooth peaks in the data,
$A_L$ can also be seen as a fudge parameter controlling the amount of
smoothing of the peaks. The Figure \ref{fig:eff} illustrates this
effect of varying $A_L$ on a concordance cosmological model.

In what follows we provide constraints on $A_L$ by analyzing a large
set of recent cosmological data.  The method we adopt is based on the
publicly available Markov Chain Monte Carlo package \texttt{cosmomc}
\cite{Lewis:2002ah} with a convergence diagnostics done through the
Gelman and Rubin statistics.  We sample the following
eight-dimensional set of cosmological parameters, adopting flat priors
on them: the baryon and cold dark matter densities $\omega_{\rm b}$ and
$\omega_{\rm c}$, the ratio of the sound horizon to the angular diameter
distance at decoupling, $\theta_s$, the scalar spectral index $n_S$,
the overall normalization of the spectrum $A$ at $k=0.002$ Mpc$^{-1}$,
the optical depth to reionization, $\tau$. Furthermore, we consider
purely adiabatic initial conditions and we impose spatial flatness.
We also consider the possibility of a massive neutrino component with
fraction $f_{\nu}>0$ and, finally, we add the weak lensing parameter $A_L$.

Our basis data set is the three--year WMAP data \cite{wmap3cosm}
(temperature and polarization) with the routine for computing the
likelihood supplied by the WMAP team. As we were approaching
completition of this paper, the five year WMAP result data became
available (\cite{wmap5cosm}, \cite{wmap5komatsu}). We have therefore
checked that our results are stable with respect to the new data.

We add the high quality and the fine-scale measurements from the ACBAR
experiment (\cite{acbar}) by using the data set provided by the team,
including normalization and beam uncertainties, window functions and
the full error covariance matrix.

Finally, we also consider an ``everything'' data set. This adds other
CMB experiments Boomerang 2K2 (\cite{boom03}), CBI (\cite{cbi}), VSAE
(\cite{vsa}), the large scale structure data in form of Red Luminous
Galaxies power spectrum (\cite{tegmark}) and the supernovae
measurements from SNLS (\cite{astier}), a prior on the Hubble's
constant from the Hubble Key project (\cite{hst}) and, finally, a Big
Bang Nucleosynthesis prior of $\omega_b=0.022\pm0.002$ at $68\%$
c.l. to help break degeneracies.

\section{Basic claim and its statistical significance}

First we run two sets of Markov-chains with $A_L$ fixed to 0 or 1. We measure
the difference between the best fit lensed model and the best fit
unlensed model of $\Delta \chi^2 = 9.34$, which is in excellent
agreement with the original claim by the ACBAR team  ($\Delta \chi^2=
9.46$).  Since both models have the same number of degrees of freedom,
this has been interpreted in \cite{acbar} as $>3\sigma$
detection of the lensing signal.

Can this difference be attributed to a single point? As can
be seen in the Table \ref{tabl:chibypoint}, where we
report the contribution to the overall $\chi^2$  coming from the 
individual points (using the full covariance information)
the answer is negative: the difference appears as randomly distributed 
across the 26 ACBAR points. 

\begin{table}
\begin{tabular}{ccc}
$\ell_{eff}$  & $\Delta \chi^2$ (lensed) & $\Delta \chi^2$ (unlensed)
  \\
\hline
$225$ & $3.3$ & $3.2$ \\
$470$ & $2.3$ & $2.0$\\
$608$ & $1.4$ & $1.4$\\
$695$ & $1.7$ & $2.4$\\
$763$ & $9.5\cdot 10^{-2}$ & $1.3\cdot 10^{-1}$\\
$823$ & $3.3\cdot 10^{-1}$ & $2.0\cdot 10^{-1}$\\
$884$ & $2.2$ & $2.3$\\
$943$ & $1.0$ & $1.8$\\
$1003$ & $2.0$ & $4.1$\\
$1062$ & $8.5\cdot 10^{-2}$ & $-1.7\cdot 10^{-2}$\\
$1122$ & $6.2\cdot 10^{-2}$ & $1.9\cdot 10^{-1}$\\
$1183$ & $6.5\cdot 10^{-2}$ & $2.2\cdot 10^{-2}$\\
$1243$ & $1.3\cdot 10^{-1}$ & $-3.6\cdot 10^{-3}$\\
$1301$ & $-3.9\cdot 10^{-3}$ & $3.1\cdot 10^{-1}$\\
$1361$ & $1.7$ & $2.3$\\
$1421$ & $1.2\cdot 10^{-1}$ & $3.4\cdot 10^{-1}$\\
$1482$ & $4.1$ & $4.9$\\
$1541$ & $1.3\cdot 10^{-1}$ & $4.5\cdot 10^{-3}$\\
$1618$ & $1.4$ & $3.6$\\
$1713$ & $1.4\cdot 10^{-2}$ & $-3.7\cdot 10^{-2}$\\
$1814$ & $3.0\cdot 10^{-1}$ & $3.2\cdot 10^{-1}$\\
$1898$ & $2.0\cdot 10^{-1}$ & $-3.5\cdot 10^{-3}$\\
$2020$ & $2.3\cdot 10^{-3}$ & $1.3\cdot 10^{-2}$\\
$2194$ & $2.7\cdot 10^{-1}$ & $5.5\cdot 10^{-1}$\\
$2391$ & $2.3$ & $2.5$\\
$2646$ & $1.1$ & $1.3$\\
\hline
total & 26.2 & 34.0 \\
\end{tabular}
\caption{\label{tabl:chibypoint} 
  This is the contribution to the overall $\chi^2$  coming from the 
  individual points, using the full covariance information.
  This quantity is not constraint to be positive, as it is equal 
  to $\Delta \chi^2_i = ((\vec{d}-\vec{t})^T
  C^{-1})_i(\vec{d}-\vec{t})_i $, where $d$ denotes data 
  vector, $t$ denotes theory vector and $C$ is the covariance matrix 
  and there is no summation over repeated indices. This table shows that 
  there are no significant outliers in the data as the overall contribution to $\chi^2$ is evenly distributed across the bins. 
  The signal is coming from a range of scales.   }
\end{table}

The effect is also marginally present in the WMAP 
third year and five year data.
Considering only the WMAP third year result we found
a $\Delta \chi^2 \sim 1.6$ between the $A_L=1$ and $A_L=0$
maximum likelihood model. Considering the newly released
WMAP five year data (\cite{wmap3cosm,wmap5komatsu}) which extend to
higher $\ell$  we get $\Delta \chi^2 \sim 3.1$.

We can ask the question of significance in the Bayesian way, which
should be more accurate in this relatively low signal-to-noise
regime. In the Bayesian theory, the relative probability of a model
(assuming the prior probabilities on each model are the same to start
with) is given by its evidence, which is the integral of likelihood
over the prior (see e.g. \cite{slosar}, \cite{liddle}).

\begin{equation}
  E = \int L(\theta) {\rm d}^N \theta
\end{equation}
As shown in \cite{marshall}, the evidence can be written
as
\begin{equation}
  \log E = \log L_{\rm max} + \left( \frac{V_L}{V_\Pi}\right),
\end{equation}
where  $L_{\rm max}$ is the likelihood at the most likely point and
$V_L$ and $V_\Pi$ are suitably defined volumes of posterior and
prior.  

The crucial point for this paper is that the evidence ratio for the
lensed and unlensed model can be written simply as 
\begin{equation}
  \Delta \log E = \Delta \log L_{\rm max} + \Delta V_L,
\end{equation}
since the prior volumes cancel exactly for the same underlying
parameter space. The posterior volume can be roughly estimated as 
\begin{equation}
  V_L \propto \prod_i \sigma_i,
\end{equation}
where $\sigma_i$ are the marginalized estimates of the errors from the
Markov Chains. A considerably better estimate would be to take
  the full error covariance into account, however, the models are so
  close that the noise in estimating the error covariance would
  probably dominate. This allows to estimate the evidence ratio to be
\begin{equation}
  E_{\rm lensed} - E_{\rm unlensed} \sim 4.67 + 0.075 = 4.75
\end{equation}

The net result is that the evidence difference is dominated by the
best-fit effect: both theories are equally good at fitting the
available parameter volume, however, the best-fit model is
considerably better for the lensed model. In fact, the volume factor
\emph{strengthens} rather than weakens the evidence for lensing in the
ACBAR data.

\begin{figure}
\centerline{\includegraphics[width=10cm]{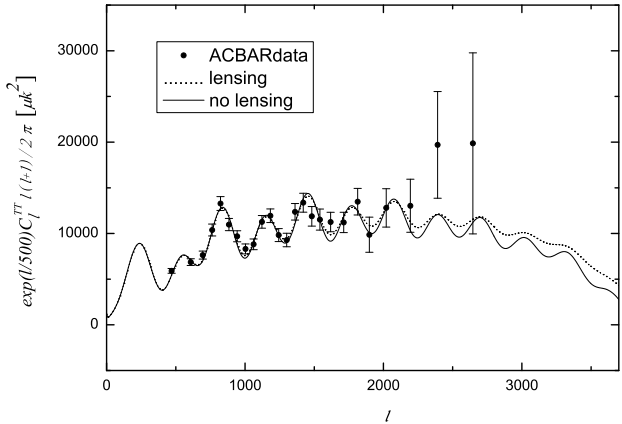}}
\caption{\label{fig:expcl}This figure shows the ACBAR data
  with $C_\ell$ spectrum predictions suitably multiplied to show the structure
  of the peaks more clearly. }
\end{figure}

\section{Varying $A_L$}

However, the anticipated forecast for the ACBAR detection from Fisher
matrix analysis is only at about $1$-sigma level.  
How are the ACBAR results at a so much higher confidence limit?

The Figure \ref{fig:expcl} show the ACBAR points plotted against
$C_\ell \ell (\ell+1) / 2\pi \exp(\ell/500)$, where the exponent has
been chosen to roughly counter-act the Silk's damping.  We see that
there is a weak ``chi-by-eye'' evidence that the ACBAR data are
actually overly-smooth given the theoretical predictions and that this
over-smoothness is driving up the detection.

\begin{figure}
\centerline{\includegraphics[width=9cm]{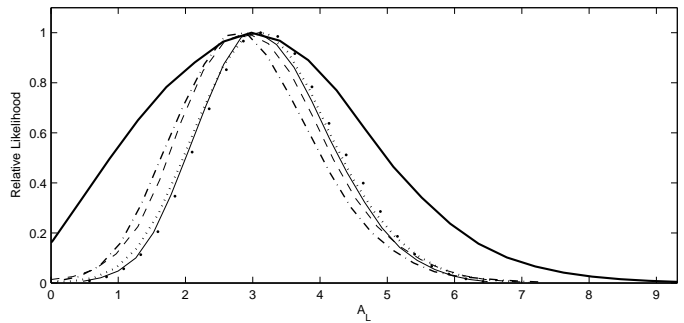}}
\caption{\label{fig:like} Marginalized $1$-D likelihood disribution
  for $A_L$ for different datasets considered: WMAP3-alone (solid
  bold), WMAP3+ACBAR (dotted), WMAP3+''everything'' (dotted bold),
  WMAP3+ACBAR+strings (solid), WMAP3+ACBAR+SZ1 (dashed),
  WMAP3+ACBAR+SZ2 (dotted-dash).}
\end{figure}

We have therefore performed additional runs where we let $A_L$ vary.
We consider the case with WMAP3 data alone, with WMAP3+ACBAR data, and
WMAP3+everything data sets. Our results are summarized in the top half of
Table \ref{tabl:1} and in Figure \ref{fig:like}.

\begin{table}
\begin{tabular}{ccc}
data set & model & limits on $A_L$\\
\hline
WMAP3 & free $A_l$ & $3.1^{+1.6+3.4}_{-1.7-2.8}$ \\
WMAP3 + ACBAR & free $A_L$ & $3.2^{+1.0+2.1}_{-0.9-1.7}$ \\
WMAP3 + everything & free $A_L$ & $3.3^{+1.0+1.9}_{-0.9-1.8}$\\
\hline
WMAP5 & free $A_l$ & $2.5^{+1.3+2.6}_{-1.2-2.1}$ \\
WMAP5 + ACBAR & free $A_L$ & $3.0^{+0.9+1.8}_{-0.9-1.6}$ \\
WMAP5 + everything & free $A_L$ & $3.1^{+0.9+1.8}_{-0.8-1.5}$\\
\hline
WMAP3 + ACBAR & +strings & $2.9^{+1.3+2.3}_{-1.2-1.8}$\\
WMAP3 + ACBAR & +SZ1 & $3.1^{+1.0+2.2}_{-1.0-2.0}$\\
WMAP3 + ACBAR & +SZ2 & $3.0^{+1.0+2.3}_{-1.0-1.8}$\\
\end{tabular}

\caption{\label{tabl:1}. This table shows results for constraints on
  the $A_L$ parameter. We report one and two sigma errors. Note that
  all results are statistically compatible with the standard
  prediction of $A_L=1$ at the level of $2$-$3$ $\sigma$.}

\end{table}

We see that the results prefer values of $A_L$ which are considerably
higher than unity. As we show below, the result is not affected by the
inclusion of the Sunyaev-Zeldovich component. Therefore, the detection
is coming from the smoothness of peaks, rather than excess of power on
the smallest scales. This can also be seen ``by eye'' in the Figure
\ref{fig:expcl}.

The level of confidence for excess is above
$2 \sigma$ (except for the WMAP data alone case which is $\sim 1
\sigma$) but less than three sigma away from one.  In agreement with a
simple Fisher matrix forecast, we find a standard deviation of the
lensing amplitude of $\Delta A_L=1$. We also looked for correlations
between $A_L$ and other parameters and found them to be negligible for
all other parameters.

Also in Table \ref{tabl:1} we report a similar analysis but now
cosidering the recent WMAP 5-year data release. As we can see, 
while the error bars are slightly reduced, the
new data confirm the results obtained with the previous WMAP 3-year 
data.

How shall we interpret these results? Let us consider three possibilities:

\begin{enumerate}

\item \textit{The result is a statistical fluctuation}. We note that
  the result is less than three sigma away from the theoretically most
  expected value of 1. The simplest explanation is that this is $2$-$3$
  $\sigma$ statistical fluctuation, with data fundamentally in agreement
  with the lensed CMB theory. However, at the same time, the unlensed
  theory is deep in the tails of the $A_L$ probability distribution
  and therefore has a considerably worse $\chi^2$. In other words,
  ACBAR had a lucky noise realization to be able to claim detection of
  lensing.

\item \textit{Hint of new physics}. It is possible that new physics is 
  responsible for over-smoothness of the
  power spectrum. This is obviously the most interesting
  option. We explore these possibilities in further detail in the
  following two sections.

\item \textit{Unknown foregrounds or 
experimental systematics}. A natural
  possibility is an unaccounted systematic in the experiment
  itself. CMB experiments are intrinsically difficult and despite many
  jack-knife tests that the authors have performed one should not
  exclude a possibility of a systematic that has slipped through.
We discuss in the next section the possibility of an unknown
foreground component.

\end{enumerate}

\section{Additional components}

We will now consider whether there could be an additional component
that could bring about smoothing. It is possible that a smooth
continuous component could lead to an effective smearing of the peaks
when the adiabatic component where reduced by an appropriate
amount. In order to check this idea we have tried to add three
different templates, whose amplitude was allowed to be free-floating:

\begin{itemize}
\item \emph{SZ template I.} A template expected from the
  Sunyaev-Zel'dovich effect as given by the analytic model of Komatsu
  and Seljak (\cite{kokatsu}).

\item \emph{SZ template II.} A similar template based on
 smoothed particle hydro-dynamics simulations \cite{zahn08a}.

\item \emph {String template}.  A template corresponding to ``wiggly
  strings'' of \cite{pogosian}. Note that the exact shape of the
  strings corresponding to a particular model is unimportant. The
  basic question we try to address is if a broad, featureless
  addition to the power spectrum can bring about a sufficient change.
\end{itemize}

Effect of these templates on the value of $A_L$ is very small as shown
in the results in the Table \ref{tabl:1} and  Figure \ref{fig:like}.
We conclude that while the data allow for some amount of extra smooth 
component, it by no means changed the ``detection'' of lensing.

\section{Non standard models}


It is certainly important to investigate if there is any possibility
 to explain the anomaly through a mechanism based on non-standard
 physics. As we pointed out in the
introduction both dark energy and modified gravity can change the
growth and amplitude of dark matter perturbations 
and thus enhance in principle the CMB weak lensing 
signal.

Dark energy could affect the growth 
by changing the expansion history and by gravitational 
feedback of the perturbations in the dark energy component
(see e.g \cite{macal}, \cite{beandore}). However 
quintessence scalar field models are generally unable 
to produce deviations larger than few percent of the 
CMB weak lensing signal. More exotic dark energy models with non-zero
anisotropic stresses (see e.g. \cite{kunz}, \cite{mota})
 could be responsible for the anomaly.

One should however consider the possibility that gravity is 
more complicated than anticipated by Einstein and 
that this modification causes more
lensing. A feature common to a broad range of modified gravity
theories is a decoupling of the perturbed Newtonian-gauge
gravitational potentials $\phi$ and $\psi$. Whereas GR predicts
$\psi=\phi$ in the presence of non-relativistic matter, a {\it
  gravitational slip}, defined as $\psi\neq\phi$, generically occurs
in modified gravity theories (see
e.g. \cite{Schimd:2004nq,Acquaviva:2004fv,Zhang:2005vt,Skordis:2005eu,
Dvali:2000hr,Lue:2005ya,Song:2006jk,Bebronne:2007qh}). 

Gravitational lensing phenomena depend
directly on the sum of the two gravitational potentials and is
strongly affected by a gravitational slip (see e.g. 
\cite{Lue:2003ky,Zhang:2007nk,Amendola:2007rr,Schmidt:2007vj,
Amin:2007wi,Jain:2007yk,Bertschinger:2008zb}).
It is therefore interesting
to investigate if $A_L > 1$ could be explained with modified
gravity and to more quantitatively connect this parameter to modified
gravity theories.

Since a very large number of models have been conceived 
here we use the parametrization of Daniel et
al. 2008 (\cite{daniel}), 
which is simple and easy to apply to several models.  
In this parameterization 
the gravitational slip is given by a function $\varpi(z)$ such that
$\psi=(1+\varpi)\phi$ and is parameterized by a single parameter
$\varpi_0$ defined as

\begin{equation}
\varpi = \varpi_0\frac{\Omega_{\Lambda}}{\Omega_m} (1+z)^{-3}.
\label{varpievolution}
\end{equation}

\noindent i.e. it starts to be relevant at dark energy (or modified
gravity) appearance.

Following \cite{daniel}, we can easily approximate the relation 
between $A_L$ and $\varpi$ as 

\begin{equation}
\label{wbar}
  A_L (\varpi) =
  \left( \frac{G_{\varpi}(z=2)}{G_{\Lambda CDM}(z=2)} \right)^2 
\left(\frac{2+\varpi}{2}\right)^2
\end{equation}

The difference in growth factors is evaluated at $z=2$, since the
lensing kernel peaks at that redshift.  Larger values of $\varpi_0$ 
correspond to larger values of $A_L$. A value of $\varpi_0\sim1.5$ could produces very
similar results on the CMB to $A_L\ge1.5$ and thus bringing the
signal inside the $1-\sigma$ cl. According to
\cite{daniel} this range of values of $\varpi_0$ is in agreement with
the measured temperature anisotropy signal on very large angular scales
but is at odds with the recent ISW detections.

\section{Systematics}

Let us in this section investigate what kind of systematic effect
could mimic the observed over-smoothing in the data.  As we have shown
in Table \ref{tabl:chibypoint} the effect is not coming for a particular rogue data point
or a small range of scales. This further constrains possible sources.

First we note that most effects that produce smoothing in real space,
such as inaccurate characterization of the beam or pointing will
induce multiplication of the real power spectrum by the Fourier
transform of the effective beam. This is unlikely to produce the
additional smoothing required to explain the hint of an
anomaly\footnote{Very contrived scenarios are possible, but these
  would imply that the Fourier transform of the effective beam
  oscillates in anti-correlation with the cosmic structure.}.

Atmospheric fluctuations could play a role. However in this case the
effect would appear as an additional smooth background component and,
as shown in Table \ref{tabl:1}, our result appears stable under this assumption.

It may however be possible that an unaccounted for systematic is
present in the data set provided by the ACBAR team, especially in the
assessment of the sky window functions. Sky coverage of the ACBAR
telescope is very complicated pattern of many fields with somewhat
fuzzy edges.  A poor characterization of the variation of noise across
the fields could, in principle, lead to to the effect observed
here. It however exceeds the scope of this paper to investigate this
thoroughly.

Finally, it is possible that the error has been induced in the final
power-spectrum estimation step of the data-reduction procedure. The
maximum-likelihood estimator employed by the ACBAR team in principle
assumes a step-wise power spectrum and the real shape of the power
spectrum has to be accounted for carefully, especially at the
signal-to-noise present in the ACBAR data.

It is clear that at the present stage systematic effects can not be
ruled out and more data is needed.  Fortunately, weak lensing will
also produce a $B$-mode polarization signal that, if observed, will
provide a fundamental cross-check.

\section{Conclusions}

We have reanalyzed the ACBAR angular power spectrum in light of the
recent detection of a lensing signal in their angular power
spectra. We tracked this down to a hint of over-smoothness in the
power spectrum, detected at $\sim 2.5 \sigma$ statistical
significance. This over-smoothness pushed the 
theory without lensing 
deep into the tails and make it a poor fit to the data.

If interpreted as real, there are several interesting possibilities.
A modified gravity can induce extra amount of lensing and we show
that a gravitational slip could bring the discrepancy to sub
$1$-$\sigma$ level. 

How does this compare with other detection of lensing in the CMB. Two
groups (\cite{smith07},\cite{hirata08}) have searched for CMB lensing
by correlating WMAP data. The WMAP data have lower intrinsic potential
for measuring CMB lensing than ACBAR, however by using more
information than the smearing of the $C_\ell$ structure (i.e. an
optimal quadratic estimator), and by correlating to galaxy surveys,
they were able to find significant evidence at the 3-$\sigma$ level.
While the mean value found is close to unity, these previous results
allow considerable freedom in overall amplitude and a reasonable fit
can be obtained with values of $A_L$ lying somewhere in between. In
particular $A_L\sim 1.7$ is compatible with both probes at less than 2
standard deviations. However a possible interpretation is that lensing
is somehow enhanced inside the ACBAR field of view, which is only
$1\%$ of that of WMAP. It will be very interesting to apply quadratic
estimator techniques using the full four-point function information to
the ACBAR maps \cite{acbar-oqe}. As the statistical error (based on
Fisher matrix forecasting) for this probe is about 4 times smaller as
compared to the smearing of acoustic peaks investigated here, we
anticipate that this will shed light on the findings of the current
paper. 

Looking at closer measurements of lensing, the weak lensing tends to
give values of $\sigma_8$ that seem only marginally higher than that
of WMAP3 (see for example \cite{Hoekstra:2005cs,Massey:2006,
  Fu:2007qq}) and consistent with the more recent WMAP5 measurements
\cite{wmap5komatsu}.  These measurements would limit the value $A_L
\lesssim 1.2$. However, the redshift spans involved are considerably
smaller with typical redshifts probed being around $\sim
0.5$. Therefore, the drastically different source redshifts imply
that these results are not in direct contradiction and that it is
conceivable that modified gravity models can be constructed that
satisfy all observational constraints.

Maybe less excitingly, but more realistically, the feature should be
interpreted as a noise realization fluctuation or explained by
unaccounted systematics. 

Future experiments as Planck, especially with the help of polarization
data, will soon shed light on this intriguing result.

\vspace{0.6cm}

\noindent {\bf Acknowledgment}\\
We thank Chao-Lin Kuo, Carlo Contaldi, Asantha Cooray and Antony Lewis for
enlightening discussions.  EC thanks University of Berkeley for
financial support and hospitality.  AS and OZ acknowledge financial
support of the Berkeley Center for Cosmological Physics.  This research
has been supported by ASI contract I/016/07/0 "COFIS".


\begin{thebibliography}{99}

\bibitem{boom}
  C.~B.~Netterfield {\it et al.}  [Boomerang Collaboration],
  Astrophys.\ J.\  {\bf 571} (2002) 604
  [arXiv:astro-ph/0104460].

\bibitem{maxima}
  R.~Stompor {\it et al.},
  Astrophys.\ J.\  {\bf 561} (2001) L7
  [arXiv:astro-ph/0105062].

\bibitem{wmap3cosm}
D. N. Spergel {\it et al.},
arXiv:astro-ph/0603449.

\bibitem{wmap5cosm}
  G.~Hinshaw {\it et al.}  [WMAP Collaboration],
  arXiv:0803.0732 [astro-ph].

\bibitem{wmap5komatsu}
  E.~Komatsu {\it et al.},
  arXiv:0803.0547 [astro-ph].

\bibitem{iswpap}
  A.~Cabre, E.~Gaztanaga, M.~Manera, P.~Fosalba and F.~Castander,
  Mon.\ Not.\ Roy.\ Astron.\ Soc.\ Lett.\  {\bf 372} (2006) L23
  [arXiv:astro-ph/0603690];
  N.~Padmanabhan, C.~M.~Hirata, U.~Seljak, D.~Schlegel, J.~Brinkmann and D.~P.~Schneider,
  Phys.\ Rev.\  D {\bf 72} (2005) 043525
  [arXiv:astro-ph/0410360];
  M.~R.~Nolta {\it et al.}  [WMAP Collaboration],
  Astrophys.\ J.\  {\bf 608} (2004) 10
  [arXiv:astro-ph/0305097];
  P.~S.~Corasaniti, T.~Giannantonio and A.~Melchiorri,
  Phys.\ Rev.\  D {\bf 71}, 123521 (2005)
  [arXiv:astro-ph/0504115].


\bibitem{isw}   S.~Ho, C.~M.~Hirata, N.~Padmanabhan, U.~Seljak and N.~Bahcall,
  arXiv:0801.0642 [astro-ph];
  T.~Giannantonio, R.~Scranton, R.~G.~Crittenden, R.~C.~Nichol, S.~P.~Boughn, A.~D.~Myers and G.~T.~Richards,
  arXiv:0801.4380 [astro-ph].

\bibitem{lensteo}
  M.~Zaldarriaga and U.~Seljak,
  Phys.\ Rev.\  D {\bf 58} (1998) 023003
  [arXiv:astro-ph/9803150];
  A.~Lewis and A.~Challinor,
  Phys.\ Rept.\  {\bf 429} (2006) 1
  [arXiv:astro-ph/0601594].

\bibitem{silk}
  J.~Silk,
  Astrophys.\ J.\  {\bf 151} (1968) 459.


\bibitem{Acquaviva:2004fv}
  V.~Acquaviva, C.~Baccigalupi and F.~Perrotta,
  Phys.\ Rev.\  D {\bf 70}, 023515 (2004)
  [arXiv:astro-ph/0403654]. 


\bibitem{acbar}
  C.~L.~Reichardt {\it et al.},
  arXiv:0801.1491 [astro-ph].


\bibitem{tegmark}
  M.~Tegmark {\it et al.}  [SDSS Collaboration],
  Phys.\ Rev.\  D {\bf 74} (2006) 123507
  [arXiv:astro-ph/0608632].

\bibitem{hst}	W.L. Freedman {\it et al.}, 
				Astrophys. J. {\bf 553}, 47 (2001).


\bibitem{cbi}
				A.~C.~S.\ Readhead {\em et al.}, 
				Astrophys.\ J.\  {\bf 609}, 498 (2004).

\bibitem{vsa}
				C.\ Dickinson {\em et al.}, 
				Mon.\ Not.\ Roy.\ Astron.\ Soc.\  {\bf 353}, 732 (2004).


\bibitem{astier}
				P.\ Astier {\em et al.}, 
				Astron.\ Astrophys.\  {\bf 447}, 31 (2006)

\bibitem{Lewis:2002ah}
A. Lewis and S. Bridle,
Phys.\ Rev.\ D {\bf 66}, 103511 (2002) (Available from
\texttt{http://cosmologist.info}.)

\bibitem{slosar} 
  A.~Slosar {\it et al.},
  Mon.\ Not.\ Roy.\ Astron.\ Soc.\  {\bf 341} (2003) L29
  [arXiv:astro-ph/0212497].


\bibitem{liddle}  D.~Parkinson, P.~Mukherjee and A.~R.~Liddle,
  Phys.\ Rev.\  D {\bf 73} (2006) 123523
  [arXiv:astro-ph/0605003].


\bibitem{marshall} 
  P.~Marshall, N.~Rajguru and A.~Slosar,
  Phys.\ Rev.\  D {\bf 73} (2006) 067302
  [arXiv:astro-ph/0412535].

\bibitem{kokatsu}
  E.~Komatsu and U.~Seljak,
  Mon.\ Not.\ Roy.\ Astron.\ Soc.\  {\bf 336}, 1256 (2002)
  [arXiv:astro-ph/0205468].

\bibitem{pogosian}
  L.~Pogosian and T.~Vachaspati,
  Phys.\ Rev.\  D {\bf 60} (1999) 083504
  [arXiv:astro-ph/9903361].


\bibitem{macal}
  C.~P.~Ma, R.~R.~Caldwell, P.~Bode and L.~M.~Wang,
  Astrophys.\ J.\  {\bf 521} (1999) L1
  [arXiv:astro-ph/9906174].


\bibitem{beandore}
  R.~Bean and O.~Dore,
  Phys.\ Rev.\  D {\bf 69} (2004) 083503
  [arXiv:astro-ph/0307100].

\bibitem{kunz}
  M.~Kunz and D.~Sapone,
  Phys.\ Rev.\ Lett.\  {\bf 98} (2007) 121301
  [arXiv:astro-ph/0612452].


\bibitem{mota}
  D.~F.~Mota, J.~R.~Kristiansen, T.~Koivisto and N.~E.~Groeneboom,
  arXiv:0708.0830 [astro-ph].

\bibitem{Schimd:2004nq}
  C.~Schimd, J.~P.~Uzan and A.~Riazuelo,
  Phys.\ Rev.\  D {\bf 71}, 083512 (2005)
  [arXiv:astro-ph/0412120].

\bibitem{Zhang:2005vt}
  P.~Zhang,
  Phys.\ Rev.\  D {\bf 73}, 123504 (2006)
  [arXiv:astro-ph/0511218].

\bibitem{Skordis:2005eu}
  C.~Skordis,
  Phys.\ Rev.\  D {\bf 74}, 103513 (2006)
  [arXiv:astro-ph/0511591].

\bibitem{Dvali:2000hr}
  G.~R.~Dvali, G.~Gabadadze, and M.~Porrati,
  Phys.\ Lett.\ B {\bf 485}, 208 (2000).
   
\bibitem{Lue:2005ya}
  A.~Lue,
  Phys.\ Rept.\  {\bf 423}, 1 (2006)
  [arXiv:astro-ph/0510068].
      
\bibitem{Song:2006jk}
  Y.~S.~Song, I.~Sawicki and W.~Hu,
  Phys.\ Rev.\  D {\bf 75}, 064003 (2007)
  [arXiv:astro-ph/0606286].
  
\bibitem{Bebronne:2007qh}
  M.~V.~Bebronne and P.~G.~Tinyakov,
  Phys.\ Rev.\  D {\bf 76}, 084011 (2007)
  [arXiv:0705.1301 [astro-ph]].



\bibitem{Lue:2003ky}
  A.~Lue, R.~Scoccimarro and G.~Starkman,
  Phys.\ Rev.\  D {\bf 69}, 044005 (2004)
  [arXiv:astro-ph/0307034].

\bibitem{Zhang:2007nk}
  P.~Zhang, M.~Liguori, R.~Bean and S.~Dodelson,
  arXiv:0704.1932 [astro-ph].
     
\bibitem{Amendola:2007rr}
  L.~Amendola, M.~Kunz and D.~Sapone,
  arXiv:0704.2421 [astro-ph].     
   
\bibitem{Schmidt:2007vj}
  F.~Schmidt, M.~Liguori and S.~Dodelson,
  Phys.\ Rev.\  D {\bf 76}, 083518 (2007)
  [arXiv:0706.1775 [astro-ph]].
     
\bibitem{Amin:2007wi}
  M.~A.~Amin, R.~V.~Wagoner and R.~D.~Blandford,
  arXiv:0708.1793 [astro-ph].

\bibitem{Jain:2007yk}
  B.~Jain and P.~Zhang,
  arXiv:0709.2375 [astro-ph].

\bibitem{Bertschinger:2008zb}
  E.~Bertschinger and P.~Zukin,
  arXiv:0801.2431 [astro-ph].
  
    
\bibitem{daniel}
  S.~F.~Daniel, R.~R.~Caldwell, A.~Cooray and A.~Melchiorri,
  arXiv:0802.1068 [astro-ph].


\bibitem{smith07}
  K.~M.~Smith, O.~Zahn and O.~Dore,
  Phys.\ Rev.\  D {\bf 76} (2007) 043510
  [arXiv:0705.3980 [astro-ph]].


\bibitem{hirata08}
  C.~M.~Hirata, S.~Ho, N.~Padmanabhan, U.~Seljak and N.~Bahcall,
  arXiv:0801.0644 [astro-ph].

\bibitem{Hoekstra:2005cs}
  H.~Hoekstra {\it et al.},
  Astrophys.\ J.\  {\bf 647}, 116 (2006)
  [arXiv:astro-ph/0511089].
 
\bibitem{Massey:2006}
  R.~Massey {\it et al.},
  Astrophys.\ J.\  {\bf 239}, 172 (2007)
  [arXiv:astro-ph/0701480]


\bibitem{Fu:2007qq}
  L.~Fu {\it et al.},
  arXiv:0712.0884 [astro-ph].
    
\bibitem{zahn08a}
  O.~Zahn {\it et al.} in preparation.
  
  \bibitem{acbar-oqe}
ACBAR team, in preparation

    
\bibitem{boom03}
  W.~C.~Jones {\it et al.},
  arXiv:astro-ph/0507494;  F.~Piacentini {\it et al.},
  arXiv:astro-ph/0507507;
  arXiv:astro-ph/0507514.

\end{thebibliography}
\end{document}